\documentclass[12pt]{article}
\usepackage{amsfonts}
\def\func#1{\mathop{\rm #1}\nolimits}%
\begin{document}

\title{A COVARIANT FORMALISM FOR\\
CHERN--SIMONS GRAVITY}
\author{Andrzej BOROWIEC \\
{\footnotesize Institute of Theoretical Physics, University of Wroc\l aw}\\
{\footnotesize pl. Maksa Borna 9, 50--204 WROC\L AW (POLAND)} \\
Marco FERRARIS and Mauro FRANCAVIGLIA \\
{{\footnotesize Dipartimento di Matematica, Universit\`{a} di Torino}}\\
{\footnotesize Via C. Alberto 10, 10123 TORINO (ITALY)}}
\maketitle

\begin{abstract}
Chern--Simons type Lagrangians in $d=3$ dimensions are analyzed
from the point of view of their covariance and globality. We use
the transgression formula to find out a new fully covariant and global
Lagrangian for Chern--Simons gravity: the price for establishing
globality is hidden in a bimetric (or biconnection) structure.
Such a formulation  allows to calculate from a global and
simpler viewpoint the energy-momentum
complex and the superpotential both for Yang--Mills and
gravitational examples.
\end{abstract}

\section{Introduction and Preliminaries}

It is well known that Einstein's gravity is trivial in dimension
$d=2$, since the curvature tensor reduces essentially to a scalar.
Also in dimension $d=3$ Einstein theory of gravitation is somehow
trivial, since the Riemann tensor reduces essentially to the
Einstein tensor. Because of this a generalization of the standard
Hilbert Lagrangian was suggested in $d=3$, by introducing, in full
analogy with gauge theories, additional terms of the Chern--Simons
type \cite{DJT}. In a previous paper of ours \cite{BFF} we have
thence tackled with the specific problem of conservation laws for
Chern--Simons type Lagrangians, both in the Yang--Mills and in the
gravitational case. In particular, we have calculated the relevant
energy--momentum complex and the superpotential for Chern--Simons
gravity in dimension $d=3$ (see also \cite{GS1} in this context).
Another technique to compute superpotentials for Chern--Simons gauge
theory which is based on
the so-called {\it cascade equation formalism} \cite{JS} has been
recently proposed in \cite{Sil}.

Let us recall that Chern--Simons Lagrangians for gravity are
non--covariant (and non--global in general) due to the presence of
cubic terms in the connection and to a non--covariant coupling of
curvature and connection, although field equations turn out to be
global and covariant. Because of this and for the sake of
simplicity, our result of \cite{BFF} were obtained in a
non--covariant framework as well as by assuming explicitly that
spacetime had a trivial topology, i.e. assuming it to be globally
diffeomorphic to an open subset of ${\Bbb R}^{3}$. The aim of the
present note is thence to provide a ``covariantised'' version of
our previous calculations, by relying on the ``background
connection method'', a covariantisation procedure which has
revealed itself to be rather useful in the case of first order
gravity \cite{FF,FF2,FFFR}. In fact, the present paper is based on
and should be considered as a direct continuation of \cite{BFF}.
In particular, we shall use the methods for computing currents and
superpotentials as presented therein.

It is known that {\em natural}, i.e. generally covariant,
Lagrangians lead to covariant Euler--Lagrange equations of motion.
The inverse statement is, in general, not true. For example, as it
was mentioned before, this holds for the following non--global (in
general) and non--invariant metric Lagrangian in dimension $d=3$:
\begin{equation}
L_{CSG}=\frac{1}{2}\varepsilon ^{\mu \nu \rho }(R_{\beta \mu \nu }^{\alpha
}\Gamma _{\alpha \rho }^{\beta }-\frac{2}{3}\Gamma _{\beta \mu }^{\alpha
}\Gamma _{\sigma \nu }^{\beta }\Gamma _{\alpha \rho }^{\sigma })
\label{eqn:LCSG}
\end{equation}
where $\Gamma _{\beta \mu }^{\alpha }$ and $R_{\beta \mu \nu
}^{\alpha }$ are the Christoffel symbols and the Riemann curvature
tensor of a metric $g_{\mu \nu }$, respectively; here $\alpha, \mu
\ldots = 1, 2, 3$. The Lagrangian (\ref{eqn:LCSG}) leads, when
varied with respect to the metric, to the following global and
covariant tensorial Euler--Lagrange equations
\begin{equation}
C^{\alpha \beta }\equiv 2\varepsilon ^{\mu \nu (\alpha } R_{\mu
;\nu }^{\beta)}=0  \label{eqn:Cotton}
\end{equation}
where $C^{\alpha \beta }$ is called the {\it York--Cotton tensor
density} and semicolon denotes metric covariant derivative. This
symmetric and traceless tensor density vanishes if and only if
$R_{\alpha\beta }$ is the Ricci curvature tensor of a locally
conformally flat metric $g$ (see \cite{CS1,CS2}). Here and above
$\varepsilon ^{\mu \nu \alpha }$ denotes the relevant
skew--symmetric Levi--Civita tensor density. It has been shown in
\cite{And} that the Lagrangian (\ref{eqn:LCSG}) is the only
obstruction to the equivariant inverse problem in $d=3$. This
Lagrangian is the gravitational counterpart of Chern--Simons
Lagrangians of gauge theories \cite{DJT,BFF}.

A similar situation occurs in fact for the case of Chern--Simons
gauge Lagrangian
\begin{equation}
L_{CS}=\frac{1}{2}\varepsilon ^{\mu \nu \rho }\func{tr}({\bf
F}_{\mu \nu}{\bf A}_{\rho }-\frac{2}{3}{\bf A}_{\mu }{\bf
A}_{\nu}{\bf A}_{\rho }) \label{eqn:LCS}
\end{equation}
where ${\bf A}_{\mu }$ is a matrix--valued gauge potential, its
curvature 2--form ${\bf F}_{\mu \nu }=\partial _{\mu }{\bf A}_{\nu
} -\partial _{\nu }{\bf A}_{\mu }+[{\bf A}_{\mu },{\bf A}_{\nu }]$
being the gauge field strength and $\func{tr}$ denoting the trace
operation for matrices (in any suitable matrix group). The
Lagrangian (\ref{eqn:LCS}) is not gauge--covariant although the
corresponding field equations ${\bf F}=0$ are. For this reason
such type of Lagrangians are sometimes called {\it quasi (or
almost) invariant}.

\section{Transgression Formula and Covariant Chern--Simons Lagrangians}

Let $G$ be any Lie group and let us denote by ${\frak g}$ the
corresponding Lie algebra. For simplicity we shall think of $G$ as
a matrix group and ${\frak g}$ as a matrix algebra with the
commutator $[\,,\,]$ as a Lie bracket. Consider a principal
$G$--bundle $P$ over a manifold $M$ (which, for the moment is
arbitrary) with a principal connection $\omega$ on $P$. Its
curvature 2--form is defined by $\Omega =d\omega +\omega \wedge
\omega $ and fulfills the Bianchi identities $D\Omega \equiv
d\,\Omega +[\omega ,\Omega ]=0$. Recall (see e.g. \cite{KN,EGH})
that $\omega $ is a ${\frak g}$--valued and $G$--equivariant
1--form that lives on the total space $P$ and which is not defined
on the base manifold $M$. Choosing a (local) section $e:
M\rightarrow P$ of $P$ we get via pull--back a (local)
matrix--valued 1--form ${\bf A}^{(e)}\equiv e^{*}\omega $ which
lives on an open domain $U\subseteq M$. This is the familiar gauge
potential (or Yang-Mills gauge field). In local coordinates
$\{x^{\mu }\}$ on $M$ it reads as ${\bf A}^{(e)}={\bf A}_{\mu
}^{(e)}dx^{\mu }$. A change of the local section $e\mapsto
e^{\prime }=e\,u$, with $u\in C^{\infty }(U,G)$, implies a
non-tensorial transformation law for the corresponding (local) gauge
potentials ${\bf A}^{(e)}\mapsto {\bf A}^{(e^{\prime})}=u^{-1}{\bf
A}^{(e)}u+u^{-1}du$; $u$ is also called a gauge transformation.
The (local) Yang--Mills field strength 2--form ${\bf F}^{(e)}\equiv
e^{*}\Omega $, however,  undergoes a tensorial transformation rule
${\bf F}^{(e)}\mapsto {\bf F}^{(e^{\prime})}=u^{-1}{\bf
F}^{(e)}u$. Because of this $\Omega$ is called a {\it tensorial}
2-form (see e.g. example 5.2, p. 76 in \cite{KN}, for the
correspondence between tensorial forms on $P$ and vector--valued
forms on the base $M$). In local coordinates we shall write ${\bf
F}^{(e)}=\frac{1}{2}{\bf F}_{\mu \nu }^{(e)}dx^{\mu }\wedge
dx^{\nu }$. On the contrary, $\omega$ is a {\it non-tensorial}
(but vertical) 1-form. By an abuse of notation from now on we
shall drop all upper indication to the section $e$.

For any two principal connection 1--forms $\omega$ and
$\bar{\omega}$ on $P$, Chern and Simons \cite{CS1,CS2} have
established the famous {\it transgression formula}
\begin{equation}
\func{tr}(\Omega \wedge \Omega )-\func{tr}(\bar{\Omega }\wedge
\bar{\Omega })=d\,[Q_{T}(\omega ,\bar{\omega})] \label{eqn:Trans}
\end{equation}
expressing the difference between two tensorial 4-forms. Here
\begin{equation}
Q_{T}(\omega ,\bar{\omega})\equiv\func{tr}(2\Omega\wedge \alpha -
d\alpha\wedge \alpha - 2\omega\wedge \alpha \wedge \alpha
+\frac{2}{3}\alpha \wedge \alpha \wedge \alpha ) \label{eqn:T3F}
\end{equation}
denotes the so--called {\it transgression 3--form} (see e.g.
\cite{EGH} p. 348), with $\alpha=\omega -\bar{\omega}$ and
$\bar{\Omega }=d\bar{\omega}+\bar{\omega}\wedge\bar{\omega}$.
Notice that $\func{tr}(\Omega\wedge\Omega)$ is a tensorial
scalar--valued 4--form on $P$. Therefore, it uniquely determines
the corresponding 4--form on the base manifold $M$, since
$e^*(\func{tr}(\Omega\wedge\Omega))=\frac{1}{4}\func{tr}({\bf
F}_{\mu \nu }{\bf F}_{\rho \sigma})dx^{\mu }\wedge dx^{\nu }\wedge
dx^{\rho }\wedge dx^{\sigma}$ does not depend on $e$. The
transgression form (\ref{eqn:T3F}) is an interesting and
intriguing object by its own. It can be easily re--expressed as
\begin{equation}
Q_{T}(\omega ,\bar{\omega})=\func{tr}(2\bar{\Omega }\wedge \alpha
+ {\bar D}\alpha \wedge \alpha +\frac{2}{3}\alpha \wedge \alpha
\wedge \alpha ) \label{eqn:T3F-2}
\end{equation}
where ${\bar D}\alpha =d\alpha +[\bar{\omega},\alpha ]$ denotes
the covariant derivative of $\alpha$ with respect to the
connection $\bar{\omega}$. Since $\alpha$ is tensorial, being the
difference of two connections, the form $Q_{T}(\omega
,\bar{\omega})$ is also a tensorial scalar--valued 3--form on $P$
which uniquely determines the corresponding 3--form on the base
manifold $M$.

The formula (\ref{eqn:Trans}) expresses a well known fact:
although the Chern 4--form $\func{tr}(\Omega \wedge \Omega )$
itself depends on the connection its cohomology class
$[\func{tr}(\Omega \wedge \Omega)]\in H^{4}(M,{\Bbb R})$ in the de
Rahm cohomology of $M$ is connection--independent since the
difference $\func{tr}(\Omega \wedge \Omega )-\func{tr}(\bar{\Omega
}\wedge \bar{\Omega })$ is exact. In more physical terms we can
also say that the Chern form, when considered as a Lagrangian,
whenever ${\rm dim} M=4$, is variationally trivial since its
variation
\begin{equation}
\delta \,\func{tr}(\Omega \wedge \Omega )=2\,d\,\func{tr}(\delta
\omega \wedge \Omega )\,
\end{equation}
is a total divergence.

From now on we shall assume that the base manifold $M$ is a
3--manifold. In this case $Q_{T}(\omega ,\bar{\omega})$ is also
closed since, of course, any 4--form on a 3--manifold vanishes
identically. Because of this it determines a cohomology class
$[Q_{T}(\omega ,\bar{\omega})]\in H^{3}(M,{\Bbb R})$ which, in
general, does not need to be trivial since $Q_{T}(\omega
,\bar{\omega})$ needs not to be exact.

In particular, by replacing $\bar{\omega}=0$ into (\ref{eqn:T3F})
one immediately recognizes the well known Chern--Simons 3--form:
\begin{equation}
P_{T}(\omega )=Q_{T}(\omega ,0)=\func{tr}(d\omega \wedge \omega +\frac{2}{3}%
\omega \wedge \omega \wedge \omega )\equiv \func{tr}(\Omega \wedge \omega -%
\frac{1}{3}\omega \wedge \omega \wedge \omega )  \label{eqn:CS3F}
\end{equation}
The Chern--Simons form (\ref{eqn:CS3F}) is also a closed,
scalar--valued, but non--tensorial 3--form, which lives on the
principal bundle $P$ and not on the base manifold. Therefore it
determines a cohomology class $[P_{T}(\omega )]\in H^{3}(P,{\Bbb
R})$ in the de Rahm cohomology of $P$, which, in general, may
depend on the connection. To see this one can use a type of
arguments similar to these presented in \cite{CS1} (Lemma 3.10).
For this purpose we calculate with a bit of algebra the following:
\begin{equation}
Q_{T}(\omega ,\bar{\omega})=P_{T}(\omega )-P_{T}(\bar{\omega})-d\func{tr}%
(\omega \wedge \bar{\omega})  \label{eqn:QT}
\end{equation}
Now it is clear that the element $[Q_{T}(\omega ,\bar{\omega})]\in
H^{3}(M,{\Bbb R})$ measures, in a certain sense, the difference
between the cohomology classes $[P_{T}(\omega )]$ and
$[P_{T}(\bar{\omega})]$. These classes are called {\it secondary
characteristic classes} for a manifold with connection.

The local Lagrangians (\ref{eqn:LCS}) can be obtained from
(\ref{eqn:CS3F}) by pull--back along local sections $e$ of $P$. If
any global section exists, i.e. if $P$ is a trivializable bundle
\footnote{If the group $G$ is simply connected then any principal
$G$--bundle over a 3--manifold is trivializable \cite{Fre}.} one
can use it to construct a global Lagrangian. In this case the
corresponding action integral
\begin{equation}
{\cal A}_{M}(\omega )=\frac{1}{8\pi\kappa }\int_{e}P_{T}(\omega
)\equiv \frac{1}{8\pi\kappa }\int_{M}e^{*}(P_{T}(\omega ))\equiv
\frac{1}{8\pi\kappa } \int_{M}L_{CS}({\bf A})  \label{eqn:08}
\end{equation}
is multivalued since its value depends on the section chosen
\cite{Fre,Oz,CF}. In fact, Chern and Simons found that this
dependence is up to a homology class of the section $e$, therefore
it must have a non--dynamical character. After introducing an
appropriate normalization constant $\kappa$ it turns out that the
actions corresponding to homologically non--equivalent sections
differ by integer values (the so--called {\it winding number}).
Alternatively, one can say that the action depends on the
connection and takes its values in the quotient ${\Bbb R}/{\Bbb
Z}$. In other words it produces a (secondary) characteristic
number for a 3--manifold with connection (see \cite{Fre} for an
exhaustive discussion).

To resume, fixing any (global) section, the Chern-Simons form
(\ref{eqn:CS3F}) pulls down to $M$ and gives the
Lagrangian (\ref{eqn:LCS}). 
This non--invariant Lagrangian produces, however, invariant and
geometrically simple equations of motion. Indeed, the variation of
(\ref {eqn:CS3F}) gives rise to the following expression (see
\cite{BFF})
\begin{equation}
\delta P_{T}=2\,\func{tr}(\delta \omega \wedge \Omega )+d\,
\func{tr}(\delta \omega \wedge \omega )  \label{eqn:dPT}
\end{equation}
which of course yields $\Omega =0$ as equation of motion. Since
$\delta \omega $ is tensorial, the Euler--Lagrange part is
tensorial too and one realizes that the whole non--invariance has
passed into the boundary term $\func{tr}(\delta \omega \wedge
\omega )$. This implies that the corresponding canonical
N\"{o}ther currents and superpotentials are not tensorial (compare
formulae (26), (27), (30) and (31) in \cite{BFF}). It means that
they are gauge (i.e., section) dependent or in other words they
live on the total space of the bundle $P$.

Our main idea in the present note is to use
\begin{equation}
L_{T}({\bf A},{\bf \bar{A}})=e^{*}(Q_{T}(\omega .\bar{\omega}))
\label{eqn:LT}
\end{equation}
as a Lagrangian 3--form on $M$. We stress again that $L_{T}$ is a
global and covariant object which lives on the base manifold $M$.
This fact is independent of the topologies of $P$, $G$ and $M$.
However, the price one has to pay for this is the bi--connection
character of the Lagrangian (\ref{eqn:LT}). We shall analyze two
cases: {\em (i)} both connections are dynamical; {\em (ii)} only
$\omega $ is dynamical while $\bar{\omega}$ is a fixed background
(non--dynamical) connection.

In terms of physically more relevant (but local) quantities
$\omega ={\bf A}_{\mu }dx^{\mu }$, $\Omega =
\frac{1}{2}{\bf F}_{\mu \nu }dx^{\mu
}\wedge dx^{\nu }$ and $\alpha ={\bf B}_{\mu }dx^{\mu }\equiv
({\bf A}_{\mu }-{\bf \bar{A}}_{\mu })dx^{\mu }$,  according to
(\ref{eqn:T3F-2}) and (\ref{eqn:QT}) one has
\begin{eqnarray}
L_{T}({\bf A},{\bf \bar{A}}) &=&\varepsilon ^{\mu \nu \rho }
\func{tr}({\bf\bar{F}}_{\mu \nu }{\bf B}_{\rho }+(\bar{D}_{\mu }
{\bf B}_{\nu }){\bf B}_{\rho }+\frac{2}{3}{\bf B}_{\mu }{\bf
B}_{\nu }{\bf B}_{\rho })  \nonumber
\\
&\equiv &\frac{1}{2}\varepsilon ^{\mu \nu \rho }\func{tr}({\bf F}_
{\mu \nu }{\bf A}_{\rho }-\frac{2}{3}{\bf A}_{\mu }{\bf A}_{\nu
}{\bf A}_{\rho }-{\bf \bar{F}}_{\mu \nu }{\bf \bar{A}}_{\rho
}+\frac{2}{3}{\bf \bar{A}}_{\mu }{\bf \bar{A}}_{\nu }{\bf
\bar{A}}_{\rho })  \nonumber \\ &&-\partial _{\mu }\,[\varepsilon
^{\mu \nu \rho }\func{tr}({\bf A}_ {\nu }{\bf \bar{A}}_{\rho })]
\label{LT}\end{eqnarray} where $dx^{\mu }\wedge dx^{\nu }\wedge
dx^{\rho }=\varepsilon ^ {\mu \nu \rho}dx^{1}\wedge dx^{2}\wedge
dx^{3}$ has been used and $\bar{D}_{\mu }\equiv\partial _{\mu
}+[{\bf \bar{A}}_{\mu },\,\cdot \,]$ denotes the directional
covariant derivative with respect to the connection
$\bar{\omega}$. Now the Lagrangian $L_{T}$ is represented by a
scalar density of weight one rather then a 3--form (see
\cite{BFF}).

The variation of (\ref{eqn:LT}) is easily  calculated from
(\ref{eqn:QT}) and (\ref{eqn:dPT}); we get:
\begin{equation}
\delta Q_{T}=2\,\func{tr}(\delta \omega \wedge \Omega
)-2\,\func{tr}(\delta \bar{\omega}\wedge \bar{\Omega
})+d\,\func{tr}((\delta \omega +\delta \bar{\omega})\wedge \alpha
)\ \label{eqn:dQT}
\end{equation}
Accordingly, (\ref{eqn:dQT}) reads now as
\begin{equation}
\delta L_{T}({\bf A},{\bf \bar{A}})=\varepsilon ^{\mu \nu \rho }
\func{tr}({\bf F}_{\mu \nu }\delta {\bf A}_{\rho }-{\bf
\bar{F}}_{\mu \nu }\delta {\bf\bar{A}}_{\rho })-\partial _{\mu }\,
[\varepsilon ^{\mu \nu \rho }\func{tr}({\bf B}_{\nu }(\delta {\bf
A}_{\rho }+\delta {\bf \bar{A}}_{\rho }))] \label{eqn:13}
\end{equation}
An infinitesimal pure gauge transformation is given by means of a
matrix--valued function (0--form) $\chi $. One has
\begin{equation}
\delta _{\chi }{\bf A}_{\mu }=D_{\mu }\chi ,\ \ \ \delta _{\chi
}{\bf \bar{A}}_{\mu }=\bar{D}_{\mu }\chi \ \ \ \mbox{and}\ \ \
\delta _{\chi }L_{T}=0
\end{equation}
i.e. $L_{T}$ is a gauge scalar. With this in mind we are able to
calculate the canonical N\"{o}ther current associated with a gauge
symmetry as
\begin{equation}
J_T^{\mu}(\chi)=\varepsilon^{\mu\nu\rho}\func{tr}[{\bf B}_{\nu
}(D_{\rho }\chi +\bar{D}_{\rho }\chi )]
\end{equation}
(compare with the calculations given in \cite {BFF}). This
quantity is weakly conserved. Due to the second N\"{o}ther
theorem, it decomposes into the so called {\it reduced current}
(which vanishes on shell) and the {\it superpotential}
\cite{BFF,JS,Sil,FF,BCJ,GS2}. The superpotential is known to represent
that part of a current which is identically conserved, does not
vanish on shell and which is enough for the computation of
conserved quantities (like charges, masses and so on). In this
case one gets explicitly (see \cite{BFF})
\begin{eqnarray}
J_T^{\mu }(\chi ) &=&2\partial _{\nu }[\varepsilon ^{\mu \nu \rho
} \func{tr}({\bf B}_{\nu }\chi )]+\varepsilon ^{\mu \nu \rho }
\func{tr}(\chi (D_{\rho }{\bf B}_{\nu }+\bar{D}_{\rho }{\bf
B}_{\nu }))  \nonumber \\ &=&\partial _{\rho }U_T^{\mu \rho
}+\varepsilon ^{\mu \nu \rho }\func{tr}(\chi {\bf F}_{\nu \rho
}-\chi {\bf \bar{F}}_{\nu \rho })
\end{eqnarray}
where the superpotential $U_T^{\mu \rho }=-U_T^{\rho \mu }$ takes
the very simple form
\begin{equation}
U_T^{\mu \rho }(\chi )=2\varepsilon ^{\mu \nu \rho }\func{tr}({\bf
B}_{\nu }\chi )
\end{equation}
The above decomposition can be easily justified by using the
identity $(D+{\bar D})\alpha =2(\Omega -\bar{\Omega })$
and by the following formula 
\begin{equation}
\func{tr}(D\chi \wedge \alpha )=\func{tr}(D(\chi
\alpha))-\func{tr}(\chi D\alpha )=d\func{tr}(\chi \alpha
)-\func{tr}(\chi D\alpha )
\end{equation}
which holds true since the trace vanishes on commutators.

A similar analysis can be performed for the diffeomorphism
invariance of $L_{T}$. Any vectorfield $\xi =\xi ^{\mu }\partial
_{\mu }$ on $M$ is just an infinitesimal diffeomorphism. Under
diffeomorphisms the gauge potentials ${\bf A}_{\mu }$ and ${\bf
\bar{A}}_{\mu }$ behave (at least locally; see the discussion
below) as 1--forms and $L_{T}$ as a scalar density of weight one.
An infinitesimal diffeomorphism transformation acts on any
(natural) geometric object over $M$ by means of the Lie derivative
${\cal L}_{\xi }$. In particular 
\begin{equation}
\delta _{\xi }{\bf A}_{\mu }\equiv {\cal L}_{\xi }{\bf A}_{\mu }=\xi
^{\alpha }\partial _{\alpha }{\bf A}_{\mu }+{\bf A}_{\alpha }
\partial _{\mu}\xi ^{\alpha },\ \ \ \delta _{\xi }L_{T}\equiv
{\cal L}_{\xi}L_{T}=\partial _{\alpha }(\xi ^{\alpha }L_{T})
\label{eqn:18}\end{equation} and similarly for $\delta _{\xi }{\bf
\bar{A}}_{\mu }\equiv {\cal L}_{\xi }{\bf \bar{A}}_{\mu }$. This
leads to the following expression for the N\"other current
\begin{equation}
J_T^{\mu }(\xi )=\xi ^{\mu }L_{T}+\varepsilon ^{\mu \nu \rho }
\func{tr}[{\bf B}_{\nu }\partial _{\alpha }({\bf A}_{\rho }+{\bf
\bar{A}}_{\rho })]\xi^{\alpha }+\varepsilon ^{\mu \nu \rho }
\func{tr}[{\bf B}_{\nu }({\bf A}_{\alpha }+{\bf \bar{A}}_{\alpha
})]\partial _{\rho }\xi ^{\alpha } \label{eqn:19}
\end{equation}
and
\begin{equation}
U_T^{\mu \rho }(\xi )=\varepsilon ^{\mu \nu \rho }\func{tr} [{\bf
B}_{\nu }({\bf A}_{\xi }+{\bf \bar{A}}_{\xi })]  \label{eqn:20}
\end{equation}
for the corresponding superpotential (compare with formulae (31)
and (32) in \cite{BFF}). Here for a simplicity we introduced the
shortcut ${\bf A}_{\xi }\equiv {\bf A}_{\alpha }\xi ^{\alpha }$.

Notice that the expressions (\ref{eqn:19}) and (\ref{eqn:20}) are
not gauge--covariant since they do contain gauge non--covariant
terms such as ${\bf A}_{\xi }$ and ${\bf \bar{A}}_{\xi }$ as well
as terms involving the partial derivatives. A similar situation is
also known in Yang--Mills theory, since the formal Lie derivative
(\ref{eqn:18}) does not fill the matrix degrees of freedom.
Strictly speaking, the group ${\rm Diff}(M)$ of all
diffeomorphisms of $M$ is not valid as a global invariance group
for the theory. The most general symmetry group is the group ${\rm
Aut}_{G}(P)$ which consists of all $G$--invariant bundle
authomorphisms $\Phi :P\rightarrow P$, i.e. the so--called
principal $G$--authomorphisms of $P$. The group ${\rm Gauge}(P)$
of all pure gauge transformations is in a natural way a subgroup
of ${\rm Aut}_{G}(P)$, while ${\rm Diff}(M)$ is not. One has
instead a surjective group homomorphism from ${\rm Aut}_{G}(P)$
onto ${\rm Diff}(M)$. The kernel of this homomorphism is of course
${\rm Gauge}(P)$.  It is clear that an infinitesimal authomorphism
$\phi $ of the principal bundle $P$ is generated by the
corresponding $G$--invariant projectable vectorfield on $P$ and it
can be
represented (at least locally) as a pair $%
\phi =(\xi ,\chi )$ with a vectorfield $\xi $ uniquely defined
\cite{BF} (see also discussion in \cite{GS1}).
Fixing some background principal connection $\omega
_{o}\equiv {\bf a}_{\mu }dx^{\mu }$ on $P$ and choosing $\chi
=-{\bf a}_{\xi }$ we may use the formula
\begin{equation}
\delta _{(\xi ,a)}{\bf A}_{\mu }\equiv {\cal L}_{\xi }{\bf A}_{\mu
} -D_{\mu }{\bf a}_{\xi }=\xi ^{\alpha }{\bf F}_{\mu \alpha }+
D_{\mu }({\bf A}_{\xi }-{\bf a}_{\xi })
\end{equation}
in order to lift the vectorfield $\xi $ on $M$ into the
corresponding $G$--invariant projectable vectorfield on $P$. Such
a lifting is not canonical, being background dependent, but it is
global. Moreover, $\omega _{o}$ is flat if and only if the
corresponding lift is a Lie algebra map. This remark generalizes a
so called {\it improved diffeomorphism} technique presented in
\cite{Jac,BH1,BH2}.

We can conclude this part by saying that the diffeomorphism
invariance of a Yang--Mills type Lagrangian is encoded into the
invariance with respect to the lifted diffeomorphisms, i.e. the
corresponding principal authomorphisms of $P$. In our case, the
total superpotential related to the diffeomorphism invariance of
the Lagrangian (\ref{eqn:T3F-2}) has the form
\begin{equation}
U_{T}^{\mu \rho }(\xi ,{\bf a})=\varepsilon ^{\mu \nu \rho }
\func{tr}[{\bf B}_{\nu }({\bf A}_{\xi }+{\bf \bar{A}}_{\xi }-2{\bf
a}_{\xi })]
\end{equation}
which is covariant but depends on the background. Finally,
choosing ${\bf a}={\bf \bar{A}}$ we find
\begin{equation}
U_{T}^{\mu \rho }(\xi ,{\bf \bar{A}})=\varepsilon ^{\mu \nu \rho }\func{tr}(%
{\bf B}_{\nu }{\bf B}_{\xi })
\end{equation}
which is fully covariant and background independent, provided
${\bf\bar A}$ is a dynamical connection.
Notice, that for a diagonal solution ${\bf A}={\bf \bar A}$ all
expressions for the superpotential automatically vanish, while the
limit ${\bf \bar A}\rightarrow 0$ reproduces the results
previously given in \cite{BFF}.

Alternatively, let us now assume that the connection
$\bar{\omega}$ is a fixed background (non--dynamical) connection.
Thus $\delta \bar{\omega}=0$ in (\ref{eqn:dQT}). The theory in
this case has only one dynamical field but the class of symmetries
is more restrictive: the gauge transformations have to keep the
background unchanged, i.e. $\delta _{\chi }{\bf \bar{A}}_{\mu
}\equiv \bar{D}_{\mu }\chi =0$. This implies $D_{\mu }\chi =[{\bf
B}_{\mu },\chi ]$ and
\begin{equation}
J_T^{\mu }(\chi )=2\varepsilon ^{\mu \nu \rho } \func{tr}({\bf
B}_{\nu }{\bf B}_{\rho }\chi )\equiv \varepsilon ^{\mu \nu \rho
}\func{tr}([{\bf B}_{\nu },{\bf B}_{\rho }]\chi )
\end{equation}
This last expression vanishes identically in the case of an
Abelian gauge group.

Recently, the so called {\it mixed Chern-Simons term} based on two
independent $U(1)$--gauge fields, one of electromagnetic origin
and the other statistical, has been successfully applied in
2--dimensional superconductivity (see \cite{AF} and references
quoted therein)\footnote{This comment is due to Ashoke Das.}.

\section{Bi--metric Chern--Simons gravity}

A particularly interesting situation appears when $P$ is the
bundle of linear frames $LM$, so that the group $G$ is the general
linear group $GL(3,{\Bbb R})$. Linear connections on $M$ are
principal connections in $LM$.

In this case $\omega $ is a ${\frak gl}(3,{\Bbb R})$--valued
1--form on the bundle $LM$ representing a linear connection on $M$
and $\Omega $ is its Riemann curvature 2--form. We can use a
coordinate section (gauge) $\{\partial _{\mu }\}$ to write down
$\omega ={\bf \Gamma }_{\mu }dx^{\mu }$ and $\Omega =\frac{1}{2}
{\bf R}_{\mu \nu }dx^{\mu }\wedge dx^{\nu }$, where $%
{\bf \Gamma }_{\mu }\equiv \Gamma _{\beta \mu }^{\alpha }$ and
${\bf R}_{\mu\nu }\equiv R_{\beta \mu \nu }^{\alpha }$ are the
standard local expressions for the connection coefficients and its
Riemann curvature tensor represented now as $3\times 3$ matrices.
Alternatively, we can also use a local (but not necessarily
coordinate) section $\{E_{i}=E_{i}^{\mu }\partial _{\mu }\}$, the
so--called {\it dreibein}. In this case, the matrix indices ${\bf
\Gamma }_{\mu }\equiv \Gamma _{j\mu }^{i}$ and ${\bf R}_{\mu \nu
}\equiv R_{j\mu \nu }^{i}$, so called ``world indices'', are
inherited from the dreibein $\{E_{i}\}$.

The Chern--Simons 3--form (\ref{eqn:CS3F}) lives then on the
bundle of linear frames $LM$ and the (local) Lagrangian
(\ref{eqn:LCSG}) can be obtained from (\ref{eqn:CS3F}) by
pull--back along a coordinate section $\{\partial _{\mu }\} $ of
$LM$. Having chosen a coordinate atlas on the base manifold, with
any coordinate neighborhood one can associate such a local
Lagrangian. On the intersection of two neighborhoods both
Lagrangians differ by a total derivative. This defines a
0--cochain of local Lagrangians in the sense of \v {C}ech
cohomology. Conservation laws for this type of non--global
Lagrangians will be investigated in detail in \cite{BFFP}. If the
manifold $M$ is parallelizable (i.e. $LM$ is a trivial bundle,
what is always the case for a compact, oriented 3--manifold), one
can also use a global (but probably no longer coordinate) dreibein
to obtain a global but not invariant Lagrangian.

Assuming that the linear connection ${\bf \Gamma }$ is the
Levi--Civita connection of some metric $g$ on $M$:
\begin{equation}
\Gamma _{\beta \mu }^{\alpha }=\frac{1}{2}g^{\alpha \sigma }
(\partial_{\beta }g_{\mu \sigma }+\partial _{\mu }g_{\sigma \beta
}-\partial _{\sigma }g_{\beta \mu })
\end{equation}
i.e. considering $g$ instead of ${\bf \Gamma }$ as the dynamical
variable, we thus obtain Chern--Simons gravity theory. The
corresponding action (\ref {eqn:08}) is metric dependent and it
produces the secondary invariant of a Riemannian manifold $(M,g)$
(see \cite{CS1,CS2}).

The transgression form (\ref{eqn:T3F-2}) gives then a new global
and bimetric Lagrangian density for a Chern--Simons gravity. The
Lagrangian (\ref{LT}) takes now the form
\begin{eqnarray}
L_{TG}(g,\bar{g}) &=&\varepsilon ^{\mu \nu \rho }\func{tr}({\bf
\bar{R}}_{\mu \nu }{\bf N}_{\rho }+(\bar{\nabla}_{\mu }{\bf
N}_{\nu }){\bf N}_{\rho }+\frac{2}{3}{\bf N}_{\mu }{\bf N}_{\nu
}{\bf N}_{\rho })  \nonumber \\
&\equiv &\frac{1}{2}\varepsilon ^{\mu \nu \rho }\func{tr}
({\bf R}_{\mu \nu }{\bf \Gamma }_{\rho }-\frac{2}{3}
{\bf \Gamma }_{\mu }{\bf \Gamma }_{\nu }%
{\bf \Gamma }_{\rho }-{\bf \bar{R}}_{\mu \nu }{\bar{{\bf \Gamma }}}_%
{\rho }+\frac{2}{3}{\bar{{\bf \Gamma }}}_{\mu }{\bar{{\bf \Gamma }}}_%
{\nu }{\bar{{\bf \Gamma }}}_{\rho })  \nonumber \\ &&-\partial
_{\mu }\,[\varepsilon ^{\mu \nu \rho }\func{tr}( {\bf \Gamma
}_{\nu }{\bar{{\bf \Gamma }}}_{\rho })]  \label{eqn:26}
\end{eqnarray}
where ${\bf N}_{\mu }\equiv {\bf \Gamma }_{\mu }-{\bar{
{\bf \Gamma }}}_{\mu} $ and $\bar{\nabla}_{\mu }\equiv \partial
_{\mu }+[{\bar{{\bf \Gamma }}}%
_{\mu },\,\cdot \,]$ denotes the Levi--Civita covariant derivative
with respect to the metric $\bar{g}$. Again, since the difference
of two connections is a tensorial 1--form $N_{j\mu }^{i}dx^{\mu }$
one plays exclusively with tensorial objects. Therefore, there is
no need to distinguish between the world and the local indices.
Accordingly, the Lagrangian density (\ref{eqn:26}) is a global and
dreibein independent 3--form on $M$. It is even fully covariant
(i.e. natural) if one considers both metrics $(g,\bar{g})$ as
dynamical fields. Now, it is well justified to use the local
expression
\begin{eqnarray}
L_{TG}(g,\bar{g}) &=&\varepsilon ^{\mu \nu \rho
}(\bar{R}_{\beta\mu\nu }^{\alpha }N_{\alpha\rho }^{\beta
}+(\bar{\nabla}_{\mu }N_{\beta \nu }^{\alpha })N_{\alpha \rho
}^{\beta }+\frac{2}{3}N_{\beta \mu }^{\alpha }N_{\sigma \nu
}^{\beta }N_{\alpha \rho }^{\sigma })  \nonumber \\ &\equiv
&\frac{1}{2}\varepsilon ^{\mu \nu \rho }(R_{\beta \mu \nu
}^{\alpha }\Gamma _{\alpha \rho }^{\beta }-\frac{2}{3}\Gamma
_{\beta \mu }^{\alpha}\Gamma _{\sigma \nu }^{\beta }\Gamma
_{\alpha \rho }^{\sigma }-\bar{R}_{\beta \mu \nu}^{\alpha
}\bar{\Gamma}_{\alpha\rho }^{\beta }+\frac{2}{3}\bar{\Gamma}_{\beta
\mu }^{\alpha}\bar{\Gamma}_{\sigma \nu }^{\beta }\bar{%
\Gamma}_{\alpha \rho }^{\sigma })  \nonumber \\ &&-\partial _{\mu
}\,[\varepsilon^{\mu\nu\rho}\Gamma_{\beta\nu }^{\alpha
}\bar{\Gamma}_{\alpha \rho }^{\beta }]  \label{eqn:27}
\end{eqnarray}
for the corresponding global 3--form on $M$. Variation of
(\ref{eqn:27}) with respect to the connections $({\bf \Gamma
},{\bar{{\bf \Gamma }}})$ yields (compare with (\ref{eqn:13})):
\begin{equation}
\delta L_{TG}=\varepsilon^{\mu \nu\rho}(R_{\beta\mu \nu}^{\alpha
}\delta \Gamma _{\alpha\rho}^{\beta}-\bar{R}_{\beta\mu\nu
}^{\alpha }\delta\bar{\Gamma}_{\alpha \rho }^{\beta})-\partial
_{\mu}\,[\varepsilon^{\mu\nu\rho}N_{\beta\nu}^{\alpha}(\delta
\Gamma_{\alpha\rho}^{\beta}+\delta\bar{\Gamma}_{\alpha\rho
}^{\beta})]  \label{eqn:28}
\end{equation}

In fact, the Lagrangian (\ref{eqn:27}) and its variation
(\ref{eqn:28}) can be alternatively analyzed from a first-order
(\'{a} la Palatini) point of view, i.e. having just two linear
connections $({\bf \Gamma },{\bar{{\bf \Gamma }}})$ as dynamical
variables \footnote{Of course, the  bi-metric and bi-connection
approaches are not equivalent since they lead to non-equivalent
equations of motion.}. As a symmetry transformation consider then
a 1--parameter group of diffeomorphisms generated by the
vectorfield $\xi =\xi ^{\alpha}\partial_{\alpha }$. The Lie
derivative of an arbitrary (non-symmetric) linear connection ${\bf
\Gamma }$ reads (see e.g. \cite{Yano,Sch})
\begin{equation}
\delta _{\xi }{\bf \Gamma }_{\rho }\equiv {\cal L}_{\xi }\Gamma
_{\alpha \rho }^{\beta }=\xi ^{\sigma }R_{\alpha \sigma \rho
}^{\beta }+\nabla _{\rho }\nabla^*_{\alpha }\xi ^{\beta }
\label{yano}
\end{equation}
where $\nabla^*_{\alpha}\xi ^{\beta}=\partial_{\alpha }\xi ^{\beta
}+\Gamma^\beta_{\sigma\alpha}\xi^\sigma$ (remember that
$\nabla_{\alpha}\xi ^{\beta}=\partial_{\alpha }\xi ^{\beta
}+\Gamma^\beta_{\alpha\sigma}\xi^\sigma$). It defines a canonical
natural lift from any vectorfield on $M$ to the corresponding
invariant projectable vectorfield on an appropriate bundle of
geometric objects over $M$. In other words, the difference between
this case and the general one discussed in the previous section is
that the Lie transport provides now a canonical (i.e. background
independent) embedding of ${\rm Diff}(M)$ into the group of
principal authomorphisms of $LM$ with a gauge part represented by
$\chi=\nabla^*\xi$. Applying formula (\ref{eqn:20}) to the present
case one might be tempted to write
\begin{equation}
U_T^{\mu \rho }(\xi )=\varepsilon ^{\mu \nu \rho }
N_{\beta\nu}^\alpha (\nabla^*_{\alpha }+\bar{\nabla}^*_{\alpha
})\xi^\beta 
\end{equation}
for the corresponding superpotential. This is wrong since
variation (\ref{yano}) is a second order differential operator in
$\xi$ (see \cite{BFF,FF,FF2}).

It is now convenient to assume that both connections are symmetric
(i.e. torsion free) linear connections on $M$. Thus following the
same steps as for computations of formula (60) in \cite{BFF} but
this time in a covariant manner, i.e. having replaced $\Gamma
_{\beta \mu }^{\alpha }$ by $N_{\beta \mu }^{\alpha }$ and the
partial derivatives $\partial _{\mu }$ by the covariant ones
$\nabla _{\mu }$ or $\bar{\nabla}_{\mu }$ respectively one gets
\begin{equation}
U_T^{\mu \rho }(\xi )=
\frac{1}{6}\varepsilon^{\mu\nu\rho}[(\nabla
_{\sigma}+\bar{\nabla}_{\sigma})(3N_{\alpha\nu}^{\sigma}
- \delta_{\nu}^{\sigma}N_{\alpha\beta}^{\beta})]\xi^{\alpha} 
- \frac{1}{3}\varepsilon^{\mu\nu\rho}(3N_{\alpha\nu}^{\sigma}
-\delta_{\nu}^{\sigma}N_{\alpha\beta}^{\beta})(\nabla_{\sigma}+
\bar{\nabla}_{\sigma })\xi ^{\alpha }\label{eqn:UT20}
\end{equation}

Coming back to the purely metric formalism we wish to perform the
variation of (\ref{eqn:27}) with respect to the metrics
$(g,\bar{g})$. For this reason one has to replace $\delta {\bf
\Gamma }_{\rho }$ in the first term of (\ref{eqn:28}) by means of
the {\em ``Palatini formula''}
\begin{equation}
\delta \Gamma _{\beta \rho }^{\alpha }=\frac{1}{2}g^{\alpha \sigma }(\nabla
_{\beta }\delta g_{\rho \sigma }+\nabla _{\rho }\delta g_{\sigma \beta
}-\nabla _{\sigma }\delta g_{\beta \rho })
\end{equation}
and the same for $\delta {\bar{{\bf \Gamma }}}_{\rho }$.
Accordingly, after some computation (see also \cite{BFF}) the
bimetric first variational formula reads now as
\begin{eqnarray}
\delta L_{TG} &=&\bar{C}^{\alpha \rho }\delta \bar{g}_{\alpha \rho
}-C^{\alpha \rho }\delta g_{\alpha \rho }+  \nonumber \\
&&\partial _{\mu }[\varepsilon ^{\mu \nu \rho }(2R_{\nu }^{\alpha
}\delta g_{\alpha \rho }-2\bar{R}_{\nu }^{\alpha }\delta
\bar{g}_{\alpha \rho }-N_{\beta \nu }^{\alpha }(\delta \Gamma
_{\alpha \rho }^{\beta }+\delta \bar{\Gamma}_{\alpha \rho }^{\beta
})]  \label{eqn:30}
\end{eqnarray}
where the York--Cotton tensor density $C^{\alpha \rho }$ (resp.
$\bar{C}^ {\alpha\rho }$) is given by (\ref{eqn:Cotton}). The
Euler--Lagrange field equations are $C^{\alpha \rho
}=\bar{C}^{\alpha \rho }=0$. We recall that the York--Cotton
tensor density is symmetric, traceless, divergence--free and it
vanishes if and only if the corresponding metric is conformally
flat.

Again, as a symmetry transformation let us consider  a flow of
diffeomorphisms generated by the vectorfield $\xi =\xi ^{\alpha }
\partial_{\alpha }$. In this case the Lie derivative operators
\begin{equation}
\delta _{\xi }g\equiv {\cal L}_{\xi }g_{\alpha\rho
}=\nabla_{\alpha}\xi _{\rho}+\nabla_{\rho }\xi_{\alpha}
\label{eqn:31}
\end{equation}
and
\begin{equation}
\delta _{\xi }{\bf \Gamma }_{\rho }\equiv {\cal L}_{\xi }\Gamma
_{\alpha \rho }^{\beta }=\xi ^{\sigma }R_{\alpha \sigma \rho
}^{\beta }+ \nabla _{\rho}\nabla _{\alpha }\xi ^{\beta }
\label{eqn:32}
\end{equation}
represent the infinitesimal variations.

Consequently, the formulae (\ref{eqn:30}), (\ref{eqn:31}) and
(\ref{eqn:32}) allow us to calculate the canonical
energy--momentum complex and superpotential in both covariant
(bi--metric) and background connection ($\delta \bar{\Gamma}\equiv
0$) formalisms. To this end we make use of the computations
already performed in \cite {BFF}. Only terms under the divergence
in (\ref{eqn:30}) will contribute into the superpotential. We see
that the first two terms correspond to the formula (56) in
\cite{BFF}. Therefore, combining with (\ref{eqn:UT20}) we arrive
to the following expression:
\begin{eqnarray}
U_{TG}^{\mu \rho }(\xi)=\varepsilon^{\mu\nu\rho}[(
3R_{\nu\alpha}-Rg_{\nu\alpha}- 3\bar{R}_{\nu\alpha}+\bar{R}
\bar{g}_{\nu\alpha})\xi^{\alpha } + \nonumber\\
\frac{1}{6}\varepsilon^{\mu\nu\rho}[(\nabla
_{\sigma}+\bar{\nabla}_{\sigma})(3N_{\alpha\nu}^{\sigma}
- \delta_{\nu}^{\sigma}N_{\alpha\beta}^{\beta})]\xi^{\alpha} 
- \frac{1}{3}\varepsilon^{\mu\nu\rho}(3N_{\alpha\nu}^{\sigma}
-\delta_{\nu}^{\sigma}N_{\alpha\beta}^{\beta})(\nabla_{\sigma}+
\bar{\nabla}_{\sigma })\xi ^{\alpha }&\ &\ \label{eqn:38}
\end{eqnarray}

As a concrete example one can consider a solution $(g, \bar g)$
consisting of a flat metric $\bar g_{\mu\nu}=\eta_{\mu\nu}$ while
$g_{\mu\nu}=\exp{(2\phi)}\,\eta_{\mu\nu}$ being conformal to
$\eta$ with a conformal factor $\phi$. Having chosen
$\xi^\alpha=\eta^{\alpha\beta}\phi_{,\beta}$ one calculates
\begin{equation}
U = d F + F d \phi
\end{equation}
as a 1--form, where
$F=-\,\eta^{\alpha\beta}\phi_{,\alpha}\phi_{,\beta}$. In
particular, for
$\phi=r\equiv\sqrt{\eta_{\alpha\beta}x^{\alpha}x^{\beta}}$ we
obtain $U=-\,d r$, i.e.
\begin{equation}
U^{\mu\rho}= -\,\varepsilon^{\mu\nu\rho}\ \frac{x_\nu}{r}
\end{equation}

\section{Conclusions}

We have considered the Chern--Simons type models in three
dimensions. Exploiting the Chern--Simons transgression $3$-form
enables us to find a new global Lagrangian density which unlike,
the local Chern-Simons Lagrangian is generally covariant. However,
in this approach the covariant Lagrangian has bi-connection
character and the corresponding theory is getting lost some of its
topological properties. Particulary, the action functional becomes
insensitive for topology of underlaying $3$-manifolds. The
formalism has been used for calculation of conserved N\"other
currents and their identically conserved parts -- superpotentials.
Two special cases are of particular interest: the case of two
connection being dynamical and the case when one of the
connections is given as a fixed background while the only the
other one is dynamical. Finally, the Chern-Simons gravity has been
treated in a similar way. In this sense the present paper
generalizes the results of our previous paper \cite{BFF} obtained
for non-covariant Chern--Simons Lagrangians (see also \cite{GS1}).
Recently, this covariant formalism has been successfully applied
to explicit numerical calculations of conserved quantities for BTZ
black hole solutions in $AdS_3$ Chern--Simons  
gravity of the Witten type \cite{Ale}.

\section*{Acknowledgments}

One of the authors (A.B.) would like to thank the INdAM--GNFM for
providing financial support and the Department of Mathematics of
Torino University for the hospitality during his stay in Italy.
The authors are grateful to Marco Godina and Marcella Palese for
fruitful discussions. Some comments by a referee are also acknowledged.


\begin{thebibliography}{99}

\bibitem{DJT}  S. Deser, R. Jackiw and S. Templeton, {\em Ann. of
Physics} {\bf 140} (1982), 372

\bibitem{BFF}  A. Borowiec, M. Ferraris and M. Francaviglia, {\em J.
Phys. A: Math. Gen. }{\bf 31} (1998), 8823 (hep-th/9801126)

\bibitem{GS1} G. Sardanashvili, {\it Energy--momentum conservation laws in gauge theory
with broken symmetry}, {\tt hep-th/0203275}

\bibitem{JS}  B. Julia and S. Silva, {\em Class. Quantum Grav.}
{\bf 15} (1998), 2173

\bibitem{Sil}  S. Silva, {\em Nucl. Phys. B }{\bf 558} (1999), 391

\bibitem{FF}  M. Ferraris and M. Francaviglia, {\em Gen. Rel. Grav.}
{\it \/}{\bf 22}(9) (1990), 965

\bibitem{FF2}  M. Ferraris and M. Francaviglia,
{\em Class. Quantum Grav. Suppl.} {\bf 9} (1992), S79

\bibitem{FFFR}  L. Fatibene, M. Ferraris, M. Francaviglia
and M. Raiteri, {\em J. Math. Phys.} {\bf 42}(3) (2001), 1173 

\bibitem{CS1}  S. S. Chern and J. Simons, {\em Proc. Nat. Acad. Sci.
USA} {\bf 68}(4) (1971), 791

\bibitem{CS2}  S. S. Chern and J. Simons, {\em Ann. Math}
{\bf 99} (1974), 48

\bibitem{And}  I. Anderson, {\em Ann. of Math.\/ }{\bf 120} (1984), 329

\bibitem{KN}  S. Kobayashi and K. Nomizu, {\it Foundations of
Differential Geometry}, John Wiley \& Sons (New York, 1963)

\bibitem{EGH}  T. Eguchi, P. B. Gilkey and A. J. Hanson,
{\em Phys. Rep.} {\bf 66} (1980), 213

\bibitem{Fre}  D. S. Freed, {\em Adv. Math.} {\bf 113} (1995), 237

\bibitem{Oz} Z. Oziewicz, {\em Rep. Math. Phys.} {\bf 31}
(1992), 85 

\bibitem{CF}  A. H. Chamseddine and J. Fr\"{o}hlich, {\em Commun. Math.
Phys.} {\bf 147} (1992), 549


\bibitem{BCJ}  D. Bak, D. Cangemi and R. Jackiw, {\em Phys. Rev.}
{\bf D 49} (1994), 5173

\bibitem{GS2} G. Sardanashvili, {\em Class. Quantum Grav.}
{\bf 14} (1997), 1371

\bibitem{BF} P.G. Bergmann and E. J. Flaherty, Jr.,
{\em J. Math. Phys.} {\bf 19} (1978), 212 

\bibitem{Jac}  R. Jackiw, {\em Phys.
Rev. Lett. }{\bf 41} (1978), 1635

\bibitem{BH1}  M. Ba\~{n}ados, L. J. Garay and M. Henneaux, {\em Phys.
Rev. D }{\bf 53}(2) (1996), R593

\bibitem{BH2}  M. Ba\~{n}ados, M. Henneaux, C. Iannuzzo and C.
M. Viallet, {\em Class. Quantum Grav.} {\bf 14} (1997), 2455

\bibitem{AF} I.J.R. Aitchison and C.D. Fosco, Phys. Rev. D {\bf
57} (1998), 1171 

\bibitem{BFFP}  A. Borowiec, M. Ferraris, M. Francaviglia and M. Palese,
{\it Conservation law for non--global Lagrangians}, Quaderni Dip.
Mat. Univ. Torino \# (2000) (submitted for publication)

\bibitem{Yano} K. Yano, {\it  The Theory of Lie Derivatives
and Its Applications}, North-Holland Publishing CO. (Amsterdam,
1957)

\bibitem{Sch} J.A. Schouten, {\it Ricci--Calculus}, (Second Ed.)
Springer--Verlag (Berlin 1954)

\bibitem{Ale} G. Allemandi, M. Francaviglia and M. Raiteri,
{\it Covariant Charges in Chern--Simons $AdS_3$ Gravity}, {\tt gr-qc/0211098},
in press in {\em Class. Quantum Grav.}

\end{thebibliography}
\end{document}